\documentclass{article}
\usepackage{amsfonts,amssymb, amsmath}
\usepackage[english]{babel}
\usepackage{hyperref}

\newtheorem{prop}{Proposition}
\newtheorem{defi}{Definition}
\newtheorem{ex}{Example}
\newenvironment{exam}{\begin{ex} \rm }{\end{ex}}

\def\bq{ \begin{equation}}
\def\eq{ \end{equation}}
\def\ben{ \begin{eqnarray}}
\def\en{ \end{eqnarray}}
\def\a{{\alpha}}

\begin{document}


\title{Superintegrable systems and Riemann-Roch theorem}

\author{A.V. Tsiganov \\
\it\small St. Petersburg State University, St. Petersburg, Russia\\
\it\small e--mail: andrey.tsiganov@gmail.com}
\date{}
\maketitle

\begin{abstract}
In algebraic geometry, there is a reduction algorithm that transforms the unreduced divisor into a unique reduced divisor, which existence is guaranteed by the Riemann-Roch theorem. We discuss  application of this algorithm to construction of finite-dimensional superintegrable systems with $n$ degrees of freedom identifying coordinates of the reduced divisor with  integrals of motion.

 \end{abstract}

\section{Introduction}
\setcounter{equation}{0}
The notion of an integrable dynamical system originates in the attempts to integrate equations of motion of the specific mechanical and astronomical problems in some explicit way. In particular, we can speak about the integrability by quadratures if we can determine all the solutions of equations of motion by means of a finite number of algebraic operations, including inversion of functions, and integrations of known functions of one variable, see \cite{koz}, page 135.

For dynamical systems integrable by Abel-Jacobi quadratures these integrals have the form
\bq\label{int-r}
\int r\bigl( x,y(x) \bigr) dx
\eq
where $r\bigl( x,y(x) \bigr)$ is a rational function on $x$ and on algebraic function $y(x)$, which satisfies equation
\bq\label{eq-curve}
\Phi\bigl(x,y(x)\bigr)=0\eq
 with an irreducible polynomial $\Phi (x,y)$ depending on the values of integrals of motion. This integral becomes well-defined upon choosing a particular branch of solutions to (\ref{eq-curve}) along a path of integration in the
 $x$-plane, that avoids the branch points where there are multiple roots. In more modern terms, one considers integral
\bq\label{int-w}
\int_\gamma \omega
\eq
on the algebraic curve $X$ defined by equation (\ref{eq-curve}) on the projective plane. Here $\omega$ is the restriction to $X$ of the rational differential 1-form $r(x, y)dx$,  path of integration $\gamma$ depends on initial conditions and avoids the singularities of $X$ and the poles of $\omega$.

 Integrals (\ref{int-r}) are highly transcendental functions of the upper limit of integration and consequently are generally difficult to study directly, but in classical mechanics we usually have the sum of such integrals
\bq\label{abel-sum }
u(t)=\sum_{i=1}^n \int_{x_0}^{x_i(t)} r\bigl( x,y(x) \bigr) dx=\sum_{i=1}^n \int_{\gamma_i} \omega\,.
\eq
The formal sum of  points $P_i (t) = (x_i, y_i) $ on $X$ is a divisor $D (t) $ of degree $n$ moving along curve $X$.

 So, when we speak of integrability by quadratures of finite-dimensional dynamical systems, we implicitly allude to replacement of  motion in physical space to motion of divisors along the algebraic curves. Evolution of divisors is governed by algebraic laws such as the B\'{e}zout theorem, Abel theorem, Riemann-Roch theorem, Brill-Noether theorem, etc. It allows us to apply these algebraic laws instead of the Emmy Noether theorem for the search of  additional conservation laws of dynamical system integrable by quadratures.

For instance, in classical mechanics divisor $D(t)$ of degree $n$ on a smooth hyperelliptic curve $X$ of genus $g$ could be
 \begin{itemize}
 \item a semi-reduced divisor at $n>g$,
 \item a reduced divisor at $n\leq g$.
 \end{itemize}
 According to the Riemann-Roch theorem each semi-reduced divisor $D$ on $X$ has a unique reduced representative $D'=\rho(D)$ of degree $g$. Movable semi-reduced divisor $D(t)$ and reduced divisor $D'$ form the intersection divisor
 \[
 D(t)+D'=0\,,\qquad \mbox{deg}D(t)=n\,,\qquad \mbox{deg}D'=g\,,
 \]
 of $X$ with  movable auxiliary curve $Y(t)$ \cite{ab}. The reduced divisor $D'=\rho(D)$ could be
 \begin{itemize}
  \item a non trivial function $D'(t)$ on time;
  \item a trivial function $D'=const$ on time.
 \end{itemize}
 If $D'$ is a constant divisor, then its coordinates are $g$ integrals of motion, which are functionally independent on $n$ integrals of motion entering into the definition of hyperelliptic curve $X$ (\ref{eq-curve}).

 In this note we want to discuss superintegrable systems having $n+g$ functionally independent integrals of motion which are associated with the Riemann-Roch theorem.

\subsection{Divisor class group}
In this section we repeat some definitions and facts from the following textbooks \cite{eh16,har77}.

Let $X$ be a hyperelliptic curve of genus $g$ defined by equation
\bq\label{h-curve}
X:\qquad y^2 + h(x)y = f(x),
 \eq
where $f(x)$ is a polynomial of degree $2g + 2$ or $2g+1$ with distinct roots, and $h(x)$ is a polynomial with deg$h\leq g$. Prime divisors are points $P_i = (x_i, y_i)$ on $X$ including points at infinity.
\begin{defi}
Divisor $D = \sum m_iP_i$, $m_i\in \mathbb Z$ is a formal sum of points on the curve, and degree of divisor $D$ is the sum $\sum m_i$ of multiplicities of points in support of the divisor.
\end{defi}
The group of divisors Div$X$ is an additive abelian group under the formal addition rule
 \[\sum m_i Z_i+\sum n_i Z_i=\sum (m_i+n_i) Z_i\,.\]
To define an equivalence relation on the divisors we use rational functions on $X$. Function $f$ is a quotient of two polynomials;  each of them is zero only on a finite closed subset of codimension one in $X$, which is therefore a union of finitely many prime divisors. The difference of these two subsets defines the principal divisor $div f$ associated with function $f$. Subgroup of Div$X$ consisting of the principal divisors is denoted by Prin$X$.
\begin{defi}
Two divisors $D, D'\in \mbox{Div} X$ are linearly equivalent
\[D\approx D'\]
if their difference $D-D'$ is principal divisor
\[
D-D'=div(f)\equiv 0\quad \mathrm{mod\, Prin}X\,.
\]
\end{defi}
 Quotient group of $\mbox{Div}X$ by  subgroup of principal divisors Prin$X$
 \[
\mbox{Pic}X =\dfrac{\mbox{Div}X}{\mbox{Prin}X}=\dfrac{\mbox{Divisors defined over k}}{\mbox{Divisors of functions defined over k}}\,.
\]
 is called the divisor class group or the Picard group.

 Restricting to   degree zero, we can also define $\mbox{Pic}^0 X= \mbox{Div}^0 X/\mbox{Prin}X$. The groups $\mbox{Pic}X$ and $\mbox{Pic}^0 X$ carry essentially the same information on $X$, since we always have
 \[
\mbox{Pic}X/\mbox{Pic}^0 X \cong \mbox{Div}X/\mbox{Div}^0 X\cong\mathbb Z\,.
 \]
 The divisor class group, where the elements are equivalence classes of degree zero divisors on $X$, is isomorphic to the Jacobian of $X$. Usually divisor and its class in Pic$X$ are denoted by the same symbol $D$.

 Some divisors on  curve $X$ are more important than others, in a sense that they contain more information about the curve itself. Indeed, in order to describe equivalence classes of degree zero divisors we can use the semi-reduced and reduced divisors.
\begin{defi}
A semi-reduced divisor is a finite sum of prime divisors of the form
\[
D=\sum m_i P_i -D_\infty\,,\qquad \mbox{deg}\,D_\infty=\sum m_i
\]
where $m_i>0$, $P_i\neq -P_j$ for $i\neq j$, no $P_i$ satisfying $P_i=-P_ i$ appears more than once, and $D_\infty$ is a linear combination of points at infinity.
\end{defi}
As semi-reduced divisors are not unique in their equivalence class we have to introduce reduced divisors.
\begin{defi}
Semi-reduced divisor $D$ is called reduced if $\sum m_i\leq g$, i.e. if the sum of multiplicities is no more than the genus of curve $C$.
\end{defi}
It follows from the Riemann-Roch theorem that for hyperelliptic curves, each equivalence class $D$ has a unique reduced representative of the form
\[
D'=\rho(D)=P_1+\cdots+P_m-D_\infty\,,\qquad \mbox{deg}D_\infty=m\,,
\]
It is enough to construction of superintegrable systems, a complete theory of divisors may be found in  \cite{eh16,har77}.

Using reduced divisors $D$ instead their equivalence classes we can describe fast and efficient algorithms for calculations  on hyperelliptic curves. Following \cite{ab,jac32}, in \cite{mum} Mumford determined polynomial representation of reduced divisor $D= (U (x), V (x)) $, i.e. a representation of the group element in $Jac (X) $:
\[U(x)=\prod(x-x_i)^{m_i}\,,\quad V(x_i)=y_i\,,\quad \mbox{deg}(V)<\mbox{deg}(U)\leq g\,,\quad v^2-f\equiv 0\,\mbox{mod}\,u\,.
\]
Here monic polynomial $U(x)$ may have multiple roots, polynomial $V(x)$ is the interpolation polynomial through all the $P_i$ according to multiplicity $m_i$.

In \cite{cant87} Cantor proposed the following algorithm for reduction of semi-reduced divisor $D$, deg$D>g$ to reduced divisor $D'$, deg$D'=g$:
\[
\begin{array}{l}
$------------------------------------------------------------------------------------------------------------$\\
\mbox{\textbf{Input}}\,\,\quad\mbox{ semi-reduced divisors}\quad D=\left(U,V\right)\quad\mbox{deg}\,D>g\\ \\
\mbox{\textbf{Output}}\quad\mbox{reduced divisor} D'=\left(U',V'\right)\quad\mbox{deg}\,D'=g\\
$------------------------------------------------------------------------------------------------------------$\\
1. \quad\mbox{while deg}(U)>g \mbox{ do}\\ \\
\phantom{1.} \quad U'\leftarrow\frac{f-hV-V^2}{U},\quad V'\leftarrow-h-V\mbox{ mod } U'\\ \\
\phantom{1.} \quad U\leftarrow\mbox{MakeMonic}(U'),\qquad V\leftarrow V'\\
 \\
2. \quad\mbox{return}\,\left(U'=U,V'=V\right)\\
$------------------------------------------------------------------------------------------------------------$
\end{array}
\]
This algorithm is a suitably formulated part of the more generic Abel's calculations \cite{ab} applicable to any algebraic curve.
In \cite{jac32}  Jacobi called polynomial $U'$ a generating function of algebraic integrals of Abel's equations.

 Below we discuss a practical application of this algorithm in classical mechanics.

\subsection{Superintegrable systems and semi-reduced divisors}
Let $X$ be a smooth projective curve of genus $g$ defined by equation
\[X:\qquad \Phi(x,y)=0\]
over an algebraically closed field. The $n$-fold symmetric product $X(n)$ of $X$ is $n$-dimensional variety, the quotient of the $n$-fold Cartesian product $X[n]$ under action of the symmetric group on $n$ letters. The points of $X(n)$ represent naturally and one-to-one the positive divisors of degree $n$ on $X$, and so in the sequel we will let
\bq\label{eff-div}
D=P_1+P_2+\ldots+P_n
\eq
either an effective divisor or the corresponding point of $X (n)$. The higher symmetric products and the corresponding semi-reduced divisors are used to construct the Jacobian of $X$, either by excision-and-glue method of Weil \cite{wei}, or the projective method of Chow \cite{matt}.

According to Jacobi \cite{jac36} variety $X (n) $ can be considered as the $n$ dimensional Lagrangian submanifold of integrable systems. Indeed, coordinates $(x_i,y_i)$, $i=1,..n$ of points $P_i$ in the support of effective divisor $D$ satisfy $n$ so-called separation relations
\bq\label{sep-rel}
 \Phi(x_i,y_i,\a_1,\cdots,\a_n)=0\,,\qquad \det\left(\dfrac{\partial^2 \Phi}{\partial x_i\partial \a_j}\right)\neq 0\,.
\eq
 where $\alpha_1,\ldots,\a_n$ are values of $n$ integrals of motion. Solving (\ref{sep-rel}) with respect to $\alpha_j$ we obtain $n$ functionally independent integrals of motion $I_j=\a_j$, which are in the involution with respect to the following Poisson brackets
 \[
 \{x_i,y_i\}=\phi_i(x_i,y_i)
 \]
 where $\phi_i$ are arbitrary functions, and other brackets are equal to zero \cite{jac36}. So, variety $X(n)$ is a  Lagrangian manifold with respect to symplectic forms associated with these brackets at $\phi_i\neq 0$. So, Jacobi join together methods of algebraic and symplectic geometry.

 Let us consider hyperelliptic curves $X$ (\ref{h-curve}) and semi-reduced divisor of zero degree
 \[
 D=P_1+P_2+\ldots+P_n-D_\infty\,,\qquad \mbox{deg}D_\infty=n\,,
 \]
 which consists of prime points $P_i(t)$ on $X$ with coordinates $(x_i,y_i)$. This semi-reduced divisor $D$ can be reduced to a unique reduced divisor $D'$, deg $D'=g$ according to the Riemann-Roch theorem. The reduced divisor $D'=\rho(D)$ could be a constant divisor or some non-trivial function on time.
\begin{exam}
 For instance, let us consider the well known addition formulae for the Legendrian elliptic integrals
\[\int\dfrac{dx_1}{(1+nx_1^2)y_1}+\int\dfrac{dx_2}{(1+nx_2^2)y_2}+
\int\dfrac{dx_3}{(1+nx_3^2)y_3}=
\ln \varphi(x_1,x_2,x_3)+const\,,
\]
 \[
\int\dfrac{(1-\kappa^2x_1^2)dx_1}{y_1}+
\int\dfrac{(1-\kappa^2x_2^2)dx_2}{y_2}+
\int\dfrac{(1-\kappa^2x_3^2)dx_3}{y_3}=\kappa^2x_1x_2x_3+const\,,
\]
 \[
\int\dfrac{dx_1}{y_1}+
\int\dfrac{dx_2}{y_2}+
\int\dfrac{dx_3}{y_3}=const\,,
\]
where an explicit form of $\phi$ may be found in \cite{barnum}. According to Abel's theorem
points $P_k=(x_k,y_k)$ on the elliptic curve
\[X: y^2=(1-x^2)(1-\kappa^2x^2)\,,\]
where
\[
x_3=\dfrac{x_1y_2+x_2y_1}{1-\kappa^2x_1^2x_2^2}\,,
\]
form an intersection divisor
\[P_1(t)+P_2(t)+P_3(t)=0\,,\]
which consists of semi-reduce divisor $D=P_1(t)+P_2(t)$ and reduced divisor $D'=P_3(t)$. Here and below for brevity we omit the points at infinity which are independent on time. The history of this geometric interpretation of Abel's results is discussed in \cite{bliss,grif04}.

Reduced divisor $D'$  consists of  point $P_3$ which is a fixed point with respect to time entering into Abel's theorem for the elliptic integrals of the first kind. For  elliptic integrals of second and third kinds point $P_3$ is a movable point.
\end{exam}

Let us formulate the main result in this note:
\begin{prop}
Let us consider integrable by quadratures dynamical system with $n$ degrees of freedom associated with the semi-reduced divisors $D(t)$ of degree $n>g$ on a hyperelliptic curve $X$ of genus $g$. If the corresponding reduced divisor $D'$ of degree $g$ is a constant divisor, then we have superintegrable system with $n+g$ functionally independent integrals of motion.
\end{prop}
Proof: Reduced divisor $D'=\rho(D)$ exists according to the Riemann-Roch theorem. The coordinates of  reduced divisor $D'$ are functionally independent on $I_j=\a_j$ according to the Abel theorem, in which these coordinates were called "algebraic constraints". Explicit expressions for these  coordinates, i.e. expressions for integrals of motion, can be obtained using standard reduction algorithm.

A set of special divisors $D$ with $\rho(D)=conts$ can be described in the framework of the Brill-Noether theory of special divisors usually formulated in sheaf cohomology terms. Below we consider some examples of such special divisors and the corresponding superintegrable systems by using the Abel \cite{ab} and Jacobi \cite{jac32} methods.

\begin{exam}
In \cite{jac32} Jacobi found algebraic integrals of Abel's equations associated with the basis of holomorphic differentials
\bq\label{jac-eqs}
\begin{array}{cccc}
 \dfrac{dx_1}{\sqrt{X_1}}&+\dfrac{dx_2}{\sqrt{X_2}}+&\cdots&+\dfrac{dx_n}{\sqrt{X_n}}=0\,,\\
 \\
 \dfrac{x_1dx_1}{\sqrt{X_1}}&+\dfrac{x_2dx_2}{\sqrt{X_2}}+&\cdots&+\dfrac{x_ndx_n}{\sqrt{X_n}}=0\,,\\
 \\
 \cdots&\cdots&\cdots&\cdots\\
 \\
 \dfrac{x_1^{n-2}dx_1}{\sqrt{X_1}}&+\dfrac{x_2^{n-2}dx_2}{\sqrt{X_2}}+&\cdots&+\dfrac{x_n^{n-2}dx_n}{\sqrt{X_n}}=0\,,
\end{array}
\eq
where $X_k=X(x_k)$ is a polynomial of $2n$-ts order on variable $x_k$.

In our terms Jacobi considere superintegrable systems with $n$ degrees of freedom associated with the semi-reduced divisor of degree $n$ on  hyperelliptic curve of genus $g=n-1$ with $h(x)=0$ and $f(x)=X(x)$ in (\ref{h-curve}). The $n-1$ equations (\ref{jac-eqs}) define the form of trajectories of these superintegrable systems, whereas the one remaining equation of motion, for instance,
\[
\dfrac{x_1^K dx_1}{\sqrt{X_1}}+\dfrac{x_2^Kdx_2}{\sqrt{X_2}}+\cdots\dfrac{x_n^K dx_n}{\sqrt{X_n}}=dt\,,\qquad K>n-2=g-1\,,
\]
defines parametrization of the trajectories by time $t$. Of course, here we can use any differential $\omega$ on the hyperelliptic curve which is independent on  holomorphic differentials $x^j/y\,dx$, $j<g-1$ entering into the system of Abel's equations (\ref{jac-eqs}).

At $n=2$ equation (\ref{jac-eqs}) coincides with the Euler differential equation on an elliptic curve. The corresponding superintegrable systems are discussed in \cite{op,ts08,ts08a,ts09}.
\end{exam}

\section{Superintegrable St\"{a}ckel systems}
Let us consider St\"{a}ckel systems with $n$ degrees of freedom defined by separated relations
\bq\label{st-eqs}
\Phi(u_j,p_{u_j})=p_{u_j}^2-f(u_j,\a_1,\ldots,\a_n)=0\,,
\eq
where $f$ is a function on $u$ and $\a_k$, so that St\"{a}ckel matrix with entries
\[
S_{ij}=\dfrac{\partial \Phi(u_j,p_{u_j})}{\partial \alpha_i}\,,\qquad i,j=1,\ldots,n,
\]
is a non-degenerate matrix. Solutions of (\ref{st-eqs}) with respect to $\a_j$ are commuting polynomials $I_j$ of second order on momenta $p_{u_i}$.

According to St\"{a}ckel \cite{st95}, Hamiltonian $H=I_1$ and Poisson brackets $\{u_i,p_{u_i}\}=1$ generate equations of motion, which can be explicitly integrated by quadratures
\bq\label{st-t}
\sum_{i=1}^n \int \dfrac{\partial \Phi(u_j,p_{u_j})}{\partial \alpha_1}\,\dfrac{du_j}{p_{u_j}}=t
\eq
and
\bq\label{st-q}
\sum_{i=1}^n \int \dfrac{\partial \Phi(u_j,p_{u_j})}{\partial \alpha_k}\,\dfrac{dx_j}{p_{u_j}}=const\,,\qquad k=2,\cdots,n\,.
\eq
The first quadrature determines parameterization of trajectories by time $t$, other quadratures determine the form of trajectories.
 According to Euler \cite{eul}, we can use these $n-1$ equations in order to describe systems with algebraic trajectories of motion.

\begin{prop}
Equations (\ref{st-q}) describe motion of the divisor $D=\sum P_i(t)$ of degree $n$  on a hyperelliptic curve $X$ (\ref{st-eqs}) of degree $g$. If $n>g$ and and this semi-reduced divisor $D$ is reduced to a constant divisor $\rho(D)$, then we have superintegrable St\"{a}ckel system with $n+g$ algebraic integrals of motion.
\end{prop}
According to Jacobi \cite{jac32}  the reduced divisor is a constant when quadratures $(\ref{st-q})$ involve a complete set of  holomorphic differentials on $X$, similar to (\ref{jac-eqs}). Below we consider only such constant reduced divisors.

\subsection{Maximally superintegrable systems}
Let us consider a case of $g=n-1$ and hyperelliptic curves of the form
\bq\label{st-jac}
\begin{array}{rcl}
X:\qquad y^2&=&\displaystyle\prod_{i=1}^M(x-e_i)\left( \sum_{j=g}^{2g+2-M} a_jx^j +\a_1x^K+\a_2x^{n-2}+\a_3x^{n-3}+\cdots+ \a_n \right)\\ \\
&=&\phi(x)\Bigl(A(x)+\a_1x^K+H(x)\Bigr)\,,
\end{array}
\eq
where $M=0,\dots, g-2$ and $K=g,\cdots 2g+2$ and $e_i,a_j\in \mathbb R$ are the parameters of our superintegrable system.
Substituting
\bq\label{subs-1}
x=u_j\qquad\mbox{and}\qquad y=\prod_{i=1}^M(u_j-e_i)p_{u_j}\,,\qquad j-1,\ldots,n,
\eq
where $u_j$ and $p_{u_j}$ are canonical coordinates $\{u_j,p_{u_j}\}=1$, into (\ref{st-jac}) and solving the resulting $n$ equations with respect to $\a_1,\ldots,\a_n$ we obtain $n$ independent polynomials $I_j=\a_j$ of second order in momenta $p_{u_k}$ commuting to each other
\[\{I_j,I_k\}=0\,.\]
These integrable systems are maximally superintegrable St\"{a}ckel systems with Hamiltonian $I_1=\alpha_1$, because for  curve $X$ (\ref{st-jac}) St\"{a}ckel's quadratures (\ref{st-q})
\[
\sum_{i=1}^n \int \dfrac{\partial \Phi(x_j,y_j)}{\partial \alpha_k}\,\dfrac{dx_j}{y}=\sum_{i=1}^n \int \dfrac{\partial H(u_j)}{\partial \alpha_k}\,\dfrac{du_j}{p_{u_i}}
=\sum_{i=1}^n \int \dfrac{u_j^{n-k}du_j}{p_{u_i}}
=const\,,\qquad k=2,\cdots,n
\]
involve all the holomorphic differentials on $X$ and, therefore, we can find $n-1$ algebraic integrals of motion directly following to \cite{jac32}.

\begin{exam}
In partial case $M=n=g+1$ we have two curves
\[
X:\qquad \Phi(x,y)= y^2-\prod_{i=1}^n (x-e_i)\cdot(a x^n+\a_1x^{n-1}+\a_2x^{n-2}+\cdots+ \a_n)=0
\]
and
\[
X:\qquad \Phi(x,y)=y^2-\prod_{i=1}^n (x-e_i)\cdot(\a_1 x^n+a x^{n-1}+\a_2x^{n-2}+\cdots+ \a_n)=0\,,
\]
which describe the $n$-dimensional harmonic oscillator
\[H=\dfrac{1}{2}\sum (p_i^2+a q_i^2)\]
and Kepler problem
\[
H=\dfrac{1}{2}\sum p_i^2+\dfrac{a}{\sqrt{\sum q_i^2}}
\]
separable in elliptic coordinates in $\mathbb R^n$:
\[
1+\sum_{i=1}^n \dfrac{q_i^2}{x-e_i}=\dfrac{U(x)}{\prod_{i=1}^n x-e_i}\,,\qquad U(x)=\prod_{j=1}^n (x-u_j)\,.
\]
Here $U(x)$ is the first Mumford coordinate of the corresponding semi-reduced divisor $D$ on $X$, whereas parameters $e_1<e_2\cdots<e_n$ determine  elliptic coordinate system
\[
u_1 < e_1 < u_2 < e_2<\cdots < u_n < e_n\,.
\]
So, this partial case corresponds to the well-known maximally superintegrable systems.
\end{exam}
\begin{exam}
In partial case $M=n-1=g$ we have three curves
\[
\begin{array}{l}
X:\qquad \Phi(x,y)=y^2-\prod_{i=1}^{n-1} (x-e_i)\cdot(a x^{n+1}+b x_n+\a_1x^{n-1}+\a_2x^{n-2}+\cdots+ \a_n)=0\,,\\
\\
X:\qquad \Phi(x,y)=y^2-\prod_{i=1}^{n-1} (x-e_i)\cdot(a x^{n+1}+ \a_1 x^n+bx^{n-1}+\a_2x^{n-2}+\cdots+ \a_n)=0\,,\\
\\
X:\qquad \Phi(x,y)=y^2-\prod_{i=1}^{n-1} (x-e_i)\cdot(\a_1 x^{n+1}+ ax^n+b x^{n-1}+\a_2x^{n-2}+\cdots+ \a_n)=0
\end{array}
\]
and three well-known superintegrable systems, separable in parabolic coordinates in $\mathbb R^n$
\[
x-2q_n-\sum_{i=1}^{n-1} \frac{q_i^2}{x-e_i}=\frac{U(x)}{\prod_{i=1}^{n-1} x-e_i}\,,\qquad U(x)=\prod_{j=1}^n (x-u_j)\,.
\]
Here $U (x) $ is the first Mumford coordinate of the corresponding semi-reduced divisor $D$ on $X$, whereas parameters $e_1<e_2\cdots<e_{n-1}$ determine local parabolic coordinate system
\[
u_1 < e_1 < u_2 < e_2<\cdots < u_n
\]
and $a,b$ are parameters of potentials in the Hamiltonian. As above, this partial case corresponds to the well-known maximally superintegrable systems.
\end{exam}
\begin{exam}
Let us present an explicit formulae for additional integrals of motion at $n=3$ and $M=2$, when hyperelliptic curve $X$ (\ref{st-jac}) is defined by equation
\[
X:\qquad y^2=x(x-a_1)(\alpha^2 x^4+\beta x^3+\a_1 x^2+\a_2 x+\alpha_3)\,.
\]
Substituting
\bq\label{subs-par}
 x= u_i\,,\qquad  y= u_i (u_i - a_1) p_{u_i}\,,\qquad i=1,2,3,
\eq
into this equation and solving the resulting equations with respect to $\a_j$ we obtain standard St\"{a}ckel's integrals of motion
\bq\label{ham-osc-3d}
\begin{array}{rcl}
I_1&=&\frac{u_1(u_1-a_1)p_{u_1}^2}{(u_1-u_3)(u_1-u_2)}
+\frac{u_2(u_2-a_1)p_{u_2}^2}{(u_3-u_2)(u_1-u_2)}
+\frac{u_3(u_3-a_1)p_{u_3}^2}{(u_2-u_3)(u_1-u_3)}\\
&-&\scriptstyle (u_1^2+u_1u_2+u_1u_3+u_2^2+u_2u_3+u_3^2)\alpha^2-(u_1+u_2+u_3)\beta\,,
\\ \\
I_2&=&\frac{u_1(a_1-u_1)(u_2+u_3)p_{u_1}^2}{(u_1-u_3)(u_1-u_2)}
+\frac{(a_1-u_2)u_2(u_1+u_3)p_{u_2}^2}{(u_3-u_2)(u_1-u_2)}
+\frac{u_3(a_1-u_3)(u_1+u_2)p_{u_3}^2}{(u_2-u_3)(u_1-u_3)}\\
&+&\scriptstyle (u_2+u_3)(u_1+u_3)(u_1+u_2)\alpha^2+(u_1u_2+u_1u_3+u_2u_3)\beta\,,
\\ \\
I_3&=& \frac{(u_1-a_1-)u_1u_3u_2p_{u_1}^2}{(u_1-u_3)(u_1-u_2)}
+\frac{(u_2-a_1)u_1u_3u_2p_{u_2}^2}{(u_3-u_2)(u_1-u_2)}
+\frac{(u_3-a_1)u_1u_3u_2p_{u_3}^2}{(u_2-u_3)(u_1-u_3)}\\
&-&\scriptstyle u_3u_2u_1(u_1+u_2+u_3)\alpha^2-u_1u_2u_3\beta\,.
\end{array}
\eq
\vskip0.2truecm
\par\noindent
\textbf{Start of the reduction algorithm}\\
Input of the reduction algorithm is a semi-reduced divisor $D=(U,V)$ with coordinates
\[
U=(x-u_1)(x-u_2)(x-u_3)
\]
and
\[\begin{array}{rcl}
V&=&\alpha x^3+b_2x^2+b_1x+b_0\\ \\
&=&(x-x_1) (x-x_2) (x-x_3) \alpha
+\frac{(x-x_2) (x-x_3) y_1}{(x_1-x_2) (x_1-x_3)}
+\frac{(x-x_1) (x-x_3) y_2}{(x_2-x_1) (x_2-x_3)}
+\frac{(x-x_1) (x-x_2) y_3}{(x_3-x_1) (x_3-x_2)}\,,
\end{array}
\]
where three coefficients $b_k$ are solutions of three algebraic equations
\[
y_k= V(x_k)\,,\qquad k=1,\ldots,3.
\]
According to \cite{ab} here we consider intersection of $X$ with a movable curve $Y(t)$ defined by equation $y=V(x)$, where polynomial $V(x)$ is defined by Lagrange interpolation at the points of the semi-reduced divisor $D$.

Substituting $y=V(x)$ into the equation $y^2-f(x)=0$ we obtain Abel's polynomial
\[\begin{array}{rcl}
\psi&=&(f-V^2)=UU'\\ \\
&=&(\beta-a_1 \alpha^2-2 \alpha b_2) x^5+(\a_1-a_1 \beta-2 \alpha b_1-b_2^2) x^4+(\a_2-\a_1 a_1-2 \alpha b_0-2 b_1 b_2) x^3\\ \\
&+&(\a_3-\a_2 a_1-2 b_0 b_2-b_1^2) x^2-(\a_3 a_1+2 b_0 b_1) x-b_0^2\,,
\end{array}
\]
where $U'$ is a polynomial from the first step of the reduction algorithm. Then we calculate polynomial
\[U=\dfrac{\psi}{U}=\dfrac{\psi}{(x-u_1)(x-u_2)(x-u_3)}=(\beta-a_1 \alpha^2-2 \alpha b_2)z^2+\cdots\]
and make it monic in order to get coordinate $U$ of the reduced divisor $D'$
\[U=x^2+A_1x+A_0\]
where
\[
\begin{array}{rcl}
A_1&=&x_1+x_2+x_3-a_1+\frac{I_1-a_1^2\alpha^2-2a_1\alpha b_2-2\alpha b_1-b_2^2}{\beta-a_1\alpha^2-2\alpha b_2}
\\ \\
A_0&=&x_1^2+x_1 x_2+x_3 x_1+x_2^2+x_2 x_3+x_3^2-\frac{a_1I_1+2\alpha b_0+2b_1b_2-I_2-(x_1+x_2+x_3)(I_1-a_1\beta-2\alpha b_1-b_2^2+I_1)}{\beta-a_1\alpha^2-2\alpha b_2}\,.
\end{array}
\]
It is enough for our purpose and, therefore, we will not calculate the second coordinate $V$.
\par\noindent
\textbf{End of the reduction algorithm.}
\vskip0.2truecm

Indeed, substituting the expressions for functions $b_k$, $I_k$ and coordinates $x_i, y_i$ (\ref{subs-par}) into $A_1, A_0$ we obtain two rational integrals of motion
\[
\{I_1,A_1\}=\{I_1,A_0\}=0\,,
\]
 which are independent on the polynomial integrals of motion $I_1,I_2$ and $I_3$. So, we get a maximally superintegrable St\"{a}ckel system.

If we identify $u_j$ in (\ref{subs-par} ) with parabolic coordinates in $\mathbb R^3$
\[
x-2q_3-\dfrac{q_1^2}{x}-\dfrac{q_2^2}{x-a_1}=\dfrac{(x-u_1)(x-u_2)(x-u_3)}{x(x-a_1)}\,,
\]
then Hamiltonian $2H=I_1$ is equal to
\[
H=\dfrac{p_1^2+p_2^2+p_3^2}{2}+\dfrac{\alpha^2}{2}(q_1^2+q_2^2+4q_3^2)+(a_1 \alpha^2-\beta)q_3+const\,.
\]
So, we obtain rational integrals of motion for oscillator separable in parabolic coordinates which  is the well known maximally superintegrable system.
\end{exam}

In a similar way we can construct other maximally superintegrable systems with $n$ degrees of freedom associated with hyperelliptic curves of genus $g$, so that $2g+3\geq n>g$. At $M\neq g,g-1$ in (\ref{st-jac}) these superintegrable systems are new.

\subsection{Non-maximally superintegrable systems}
If $g=n-2$, we can take one of the hyperelliptic curves of the form
\bq\label{st-jac2}
\begin{array}{rcl}
X:\qquad y^2&=&\displaystyle\prod_{i=1}^M(x-e_i)\left( \sum_{j=g-1}^{2g+2-M} a_jx^j +\a_1x^K+\a_2x^{L}+\a_3x^{n-3}+\cdots+ \a_n \right)\\ \\
&=&\phi(x)\left(A(x)+\alpha_1x^K+\alpha_2X^m+H(x)\right)\,.
\end{array}
\eq
In this case we have a semi-reduced divisor $D$ of degree $n$ on hyperelliptic curve $X$ of genus $g=n-2$. Substituting
\[
x=u_j\qquad\mbox{and}\qquad y=\prod_{i=1}^M(u_j-e_i)p_{u_j}\,,
\]
where $u_j$ and $p_{u_j}$ are canonical coordinates $\{u_j,p_{u_j}\}=1$ into $ (\ref{st-jac2})$ and solving the resulting $n$ equations with respect to $\a_1,\ldots,\a_n$ we obtain $n$ independent polynomials $I_j=\a_j$ of second order in momenta $p_{u_k}$ commuting to each other
\[\{I_j,I_k\}=0\,.\]
These integrable systems are superintegrable St\"{a}ckel systems with respect to Hamiltonians $I_1=\alpha_1$ or $I_2=\alpha_2$, because for curve $X$ (\ref{st-jac2}) St\"{a}ckel's quadratures (\ref{st-q})
\[
\sum_{i=1}^n \int \dfrac{\partial \Phi(x_j,y_j)}{\partial \alpha_k}\,\dfrac{dx_j}{y}=\sum_{i=1}^n \int \dfrac{\partial H(u_j)}{\partial \alpha_k}\,\dfrac{du_j}{p_{u_i}}
=\sum_{i=1}^n \int \dfrac{u_j^{n-k}du_j}{p_{u_i}}
=const\,,\qquad k=3,\cdots,n
\]
involve all the holomorphic differentials on $X$ and, therefore, we can find algebraic integrals of motion directly following to \cite{jac32}.

\begin{exam}
For completeness, we present one partial example of such systems at $n=3$, which was discussed in \cite{ts19k}.

Let us consider symmetric product $X[3]$ of the elliptic curve $X$ defined by equation
\bq\label{ell-3d}
X:\quad \Phi(x,y)=y^2-f(x)=0\,,\qquad f(x)=\alpha x^4+\beta x^3+\a_1x^2+\a_2x+\a_3\,,
\eq
and identify coordinate of point in $X[3]$, i.e. affine coordinates of the corresponding effective divisor $D$ (\ref{eff-div}), with canonical coordinates in $T^*\mathbb R^3$
\bq\label{trans-44}
x_1=u_1\,,\quad y_1=p_{u_1}\,,\quad x_2=u_2\,,\quad y_2=p_{u_2}\,,\quad x_3=u_3\,,\quad y_3=p_{u_3}\,.
\eq
According to Jacobi it allows us to joint algebraic and symplectic methods of investigations.

Solving equations $\Phi(x_i,y_i)=0$ w.r.t $\a_k$ we obtain three independent integrals of motion
\bq
\label{int-3d}
\begin{array}{rcl}
I_1&=&\scriptstyle \frac{p_{u_1}^2}{(u_1-u_3)(u_1-u_2)}+\frac{p_{u_2}^2}{(u_2-u_3)(u_2-u_1)}
+\frac{p_{u_3}^2}{(u_3-u_1)(u_3-u_2)}- (u_1^2+u_2^2+u_3^2+u_1u_2+u_1u_3+u_2u_3)\alpha-(u_1+u_2+u_3)\beta\\ \\
I_2&=&
\scriptstyle -\frac{(u_2+u_3)p_{u_1}^2}{(u_1-u_3)(u_1-u_2)}-\frac{(u_1+u_3)p_{u_2}^2}{(u_2-u_3)(u_2-u_1)}
-\frac{(u_1+u_2)p_{u_3}^2}{(u_3-u_1)(u_3-u_2)}
+(u_1+u_2)(u_1+u_3)(u_2+u_3)\alpha+
(u_1u_2+u_1u_3+u_2u_3)\beta\\ \\
I_3&=&\scriptstyle
\frac{u_2u_3p_{u_1}^2}{(u_1-u_3)(u_1-u_2)}+\frac{u_1u_3p_{u_2}^2}{(u_2-u_3)(u_2-u_1)}
+\frac{u_1u_2p_{u_3}^2}{(u_3-u_1)(u_3-u_2)}
-u_1u_2u_3(u_1+u_2+u_3)\alpha -u_1u_2u_3\beta\,.
\end{array}
\eq
The corresponding St\"{a}ckel quadratures
\[\begin{array}{rcl}
\displaystyle \int \dfrac{u_1^2du_1}{\sqrt{f(u_1)}}+\int \dfrac{u_2^2du_2}{\sqrt{f(u_2)}}+\int \dfrac{u_3^2du_3}{\sqrt{f(u_3)}}=t\,,\\ \\
\displaystyle \int \dfrac{u_1du_1}{\sqrt{f(u_1)}}+\int \dfrac{u_2du_2}{\sqrt{f(u_2)}}+\int \dfrac{u_3du_3}{\sqrt{f(u_3)}}=const\,,\\ \\
\displaystyle \int \dfrac{du_1}{\sqrt{f(u_1)}}+\int \dfrac{du_2}{\sqrt{f(u_2)}}+\int \dfrac{du_3}{\sqrt{f(u_3)}}=const\,.
\end{array}
\]
involve a  holomorphic differential on elliptic curve and, therefore, the arithmetic equation for intersection divisor
\bq\label{div-eq-3}
\Bigl(P_1(t)+P_2(t)+P_3(t)\Bigr)+P_4=D(t)+D'=0\,.
\eq
is a sum of movable divisor $D(t)$ of third degree  and constant divisor $D'$ of first degree.

\vskip0.2truecm
\par\noindent
\textbf{Start of the reduction algorithm}\\
Algorithm input consists of coordinates of the semi-reduced divisor $D(t)$ of third degree 
\[
U=(x-x_1)(x-x_2)(x-x_3)\,,\qquad V=b_2x^2+b_1x+b_0
\]
 where $b_2,b_1$ and $b_0$ are solution of the equations
\[
 y_1=b_2x_1^2+b_1x_1+b_0\,,\quad y_2=b_2x_2^2+b_1x_2+b_0\,,\quad y_3=b_2x_3^2+b_1x_3+b_0\,.
\]
Substituting coordinates of the semi-reduced divisor $D$ into the polynomial
\[U'=\frac{f-V^2}{U}\]
and making it monic, we obtain coordinates $U'=(x-x_4)$ and $V'=y_4$ of the reduced divisor $D'$, where
\[
x_4= -x_1-x_2-x_3-\dfrac{a_3-2b_1b_2}{a_4-b_2^2}\,,\qquad y_4=-V(x_4)\,.
\]
\par\noindent
\textbf{End of the reduction algorithm.}
\vskip0.2truecm
After substitution (\ref{trans-44}) we obtain rational functions $x_4$ and $y_4$ on  phase space $T^*\mathbb R^3$
\bq\label{coord-p4}
x_4=-u_1-u_2-u_3-\dfrac{\beta-2b_1b_2}{\alpha-b_2^2}\,,\qquad y_4=-(b_2x_4^2+b_1x_4+b_0)\,,
\eq
where
\[\begin{array}{rcl}
b_2&=&\dfrac{(u_2-u_3)p_{u_1}+(u_3-u_1)p_{u_2}+(u_1-u_2)p_{u_3}}{(u_1-u_2)(u_1-u_2)(u_2-u_3)}\,,\\ \\
b_1&=&-\dfrac{(u_2^2-u_3^2)p_{u_1}-(u_3^2-u_1^2)p_{u_2}-(u_1^2-u_2^2)p_{u_3}}{(u_1-u_2)(u_1-u_2)(u_2-u_3)}\,,\\ \\
b_0&=&\dfrac{u_2u_3(u_2-u_3)p_{u_1}+u_1u_3(u_3-u_1)p_{u_2}+u_1u_2(u_1-u_2)p_{u_3}}{(u_1-u_2)(u_1-u_2)(u_2-u_3)}\,.
\end{array}
\]
These functions are the first integrals of the dynamical system determined by Hamiltonians $I_1$ or $I_2$ associated with nonholomorphic  differentials on $X$
\[
\{I_1,x_4\}=\{I_1,y_4\}=0\quad\mbox{and}\quad \{I_2,x_4\}=\{I_2,y_4\}=0\,.\]
Complete algebra of the first integrals is discusses in \cite{ts19k}.
\end{exam}

\section{Effective divisors with multiple points}
In Section 2 we consider the $n$-fold symmetric product $X(n)$ of hyperelliptic curve $X$ and effective divisors $D$ (\ref{eff-div}) of the form
\[
D=P_1+P_2+\ldots+P_n\,.
\]
If we take generic positive divisors
\[
D=m_1P_1+m_2P_2+\cdots+m_nP_n\,,\qquad m_i\in\mathbb Z_+
\]
associated with sums of integrals of the holomorphic differentials on $X$
\[
m_1 \int \dfrac{ x_1^kdx_1}{y_1}+m_2 \int \dfrac{ x_2^kdx_2}{y_2}+\cdots+m_n \int \dfrac{ x_n^kdx_2}{y_n}=
\int \dfrac{ x_1^kdx_1}{y_1/m_1}+\int \dfrac{ x_2^kdx_2}{y_2/m_2}+\cdots+ \int \dfrac{ x_n^kdx_2}{y_n/m_n}
\,,
\]
we can "restore" the symmetry between points by using substitution
\[
 \dfrac{y_i}{m_i}\to y_i\,,\qquad i=1,\ldots,n,
\]
that corresponds to a noncanonical transformation of variables in phase space of the St\"{a}ckel integrable systems  (\ref{subs-1})
\bq\label{n-trans}
p_{u_i}\to \dfrac{p_{u_i}}{m_i}\,,\qquad i=1,\ldots,n.
\eq
So, all the superintegrable systems described in Section 2 remain superintegrable  after the
noncanonical transformations (\ref{n-trans}), see discussion in \cite{ts19st,ts19s,ts19k}.

\begin{exam}
Let us take  3D oscillator from  Example 5 and make the noncanonical transformation (\ref{n-trans}) of the corresponding momenta
\[
p_{u_1}\to \dfrac{p_{u_1}}{2}\,,\qquad p_{u_2}\to p_{u_2}\,,\qquad p_{u_3}\to p_{u_3}\,.
\]
This transformation changes metric in Hamiltonian (\ref{ham-osc-3d})
\[
\begin{array}{rcl}
I_1&=&\dfrac{u_1(u_1-a_1)}{(u_1-u_3)(u_1-u_2)}\dfrac{p_{u_1}^2}{4}
+\dfrac{u_2(u_2-a_1)}{(u_3-u_2)(u_1-u_2)}p_{u_2}^2
+\dfrac{u_3(u_3-a_1)}{(u_2-u_3)(u_1-u_3)}p_{u_3}^2\\ \\
&-& (u_1^2+u_1u_2+u_1u_3+u_2^2+u_2u_3+u_3^2)\alpha^2-(u_1+u_2+u_3)\beta\,,
\end{array}
\]
and doubles the first point in the movable effective divisor
\[
D=2P_1(t)+P_2(t)+P_3(t).
\]
\vskip0.2truecm
\par\noindent
\textbf{Start of the reduction alghorithm}\\
Input of the reduction algorithm is a semi-reduced divisor $D=(U,V)$ with coordinates
\[
U=(x-u_1)^2(x-u_2)(x-u_3)
\]
and
\[V=b_3 x^3+b_2x^2+b_1x+b_0\]
where four coefficients $b_k$ are solutions of the four algebraic equations
\[
y_k= V(x_k)\,,\quad k=1..3,\quad\mbox{and}\quad \dfrac{dV}{dx}( x_1) =\dfrac{1}{2y_1}\dfrac{df}{dx}(x_1)\,.
\]
due to the standard Hermite interpolation procedure. In order to get the first coordinate of the reduced divisor $D'$ we have to calculate polynomial
\[U'=\dfrac{f-V^2}{U}=(\alpha^2-b_3^2)\,(x^2+A_1x+A_0)\]
and make it monic
\[
U=x^2+A_1x+A_0\,.
\]
Coefficients of this polynomial are equal to
\[
A_1=2x_1 + x_2 + x_3 -\frac{(a_1\alpha^2 + 2b_2b_3 -\ beta}{\alpha^2 - b_3^2}
\]
and
\[\begin{array}{rcl}
A_0&=&\dfrac{2x_1^2+x_2^2+x_3^2+A_1^2-a_1^2}{2}+\dfrac{I_1}{\alpha^2 - b_3^2}\\ \\
&-&
\dfrac{\alpha^2\bigl(2a_1^2b_3^2 + 4(a_1b_2 + b_1)b_3 + 2b_2^2\bigr)
+\beta^2 - 2b_3(a_1b_3 + 2b_2)\beta - b_3^2(a_1^2b_3^2 + 4b_1b_3 - 2b_2^2)}{(\alpha^2 - b_3^2)^2}
\end{array}
\]
It is enough for our purpose and, therefore, we omit for brevity calculation of the second coordinate of $D'$ 
\par\noindent
\textbf{End of the reduction algorithm.}
\vskip0.2truecm

Substituting expressions for functions $b_k$, $I_k$ and coordinates $x_i,y_i$ (\ref{subs-par}) into $A_1,A_0$ we obtain two rational integrals of motion
\[
\{I_1,A_1\}=\{I_1,A_0\}=0\,,
\]
 which are independent of the St\"{a}ckel integrals of motion $I_1,I_2$ and $I_3$. Complete algebra of polynomial $I_1,I_2,I_3$ and rational $A_0,A_1$ integrals of motion are the polynomial algebra similar to the algebras from \cite{ts19k}.
\end{exam}

\begin{exam} In order to present calculations with a few rounds in a do - loop, we
 consider a 2D oscillator separable in parabolic coordinates on the plane
\[
u_1 = q_2-\sqrt{q_1^2+q_2^2},\qquad u_2 = q_2+\sqrt{q_1^2+q_2^2}\,,\qquad u_1<0<u_2\,.
\]
Noncanonical transformation (\ref{n-trans}) of the corresponding momenta
\[
p_{u_1}=\dfrac{p_2}{2}-\dfrac{q_2+\sqrt{q_1^2+q_2^2}}{2q_1}p_1\,,\qquad
p_{u_2}=\dfrac{p_2}{2}-\dfrac{q_2-\sqrt{q_1^2+q_2^2}}{2q_1}p_1\,
\]
transforms original Hamiltonian 
\[
H=2I_1=\dfrac{p_1^2+p_2^2}{2}-2\alpha^2(q_1^2+4q_2^2)
\]
to the Hamiltonian
\bq\label{h31}
H_{m_1m_2}=\dfrac{(m_1^2+m_2^2)(p_1^2+p_2^2)}{4 m_1^2 m_2^2}
-\dfrac{(m_1^2-m_2^2)(q_2p_1^2-2q_1p_1p_2-q_2p_2^2)}{4m_1^2m_2^2\sqrt{q_1^2+q_2^2}}
-2\alpha^2(q_1^2+4q_2^2)\,,
\eq
associated with an effective divisor $D=m_1P_1+m_2P_2$ on elliptic curve
\[E:\quad \Phi(x,y)=y^2-f(x)=0\,,\quad
f(x)=x\Bigl(\alpha^2x^3+I_1 x+I_2\Bigr)\,,
\]
It follows from the Riemann-Roch theorem that Hamiltonian $H_m$ remains superintegrable, i.e. it is in involution with two independent integrals of motion. The first of them is given by $I_2$, the second independent integral of motion is given by coordinates of the constant reduced divisor $P_3$ in the equation
\[
m_1P_1(t)+m_2P_2(t)+P_3=0\,.
\]
Below we show the reduction algorithm at $m_1=3$ and $m_2=1$.

\vskip0.2truecm
\par\noindent
\textbf{Start of the reduction alghorithm}\\
Input of the reduction algorithm is the following coordinates of the semi-reduced divisor $D$ of fourth degree:
\[
U(x)=(x-x_1)^3(x-x_1)\,,\quad V(x)=b_3x^3 + b_2x^2 + b_1x + b_0.
\]
Because polynomial $V(x)$ is a Hermite interpolation polynomial, its four coefficients $b_k$ are solutions of the  four equations
\bq\label{b-coeff}
y_1=V(x_1)\,,\quad \dfrac{dV}{dx}(x_1)=\dfrac{1}{2y_1}\dfrac{df}{dx}(x_1)\,,\quad \dfrac{d^2V}{dx^2}(x_1)= \dfrac{1}{2y_1}\dfrac{d^2f}{dx^2} - \dfrac{1}{4y_1^3}\left(\dfrac{df}{dx}(x_1)\right)^2\,.\quad
y_2=V(x_2)\,,
\eq
We start the first round of the do-loop  with calculation of coordinates of the semi-reduced divisor of second degree. We calculate
\bq\label{A-coeff}
\begin{array}{rcl}
U'&=&\dfrac{f-V^2}{U}=-b_3^2 x^2-b_3(3b_3x_1 + b_3x_2 + 2b_2)x\\ \\
&-&
b_3^2(6x_1^2 + 3x_1x_2+ x_2^2) +-b_3(6b_2x_1+-2b_2x_2 + 2b+1) + \alpha^2 - b^2_2
\end{array}
\eq
and make it monic
\[
U=\dfrac{U'}{-b_3^2}=x^2+A_1x+A_0\,.
\]
Then we have to calculate second coordinate of the semi-reduced divisor of second degree 
\[
V'=-V\, \mbox{mod}\, U'=-(b_3x^3 + b_2x^2+ b_1x+ b_0 \,\mbox{mod}\, (-b_3^2 x^2-b_3^2xA_1-b_3^2A_2)\,.
\]
Using standard modular arithmetic with polynomials we obtain
\[
V'=\alpha^2x^2+B_1x+B_0\,,
\]
where
\bq\label{B-coeff}
B_1=-b_3A_1^2 + (\alpha + b_2)A_1 +b_3 A_0 - b_1\,,\quad B_1=(\alpha+b_2-b_3A_1)A_0 - b_0,
\eq
and put $V=V'$.

At the second round of  the do-loop, we use coordinates of semi-reduced divisor of second degree 
\[U=x^2+A_1x+A_0\,,\qquad V=\alpha^2x^2+B_1x+B_0\]
in order to obtain polynomial
\[
U'=\dfrac{f-V^2}{U}=-2\alpha B_1 (z-K)
\]
and make it monic
\bq\label{x3-coord}
U=z-x_3\,,\qquad x_3=\dfrac{\alpha(2B_0-2A_1B_1) + B_1^2 - I_1)}{2\alpha B_1}\,.
\eq
As a result, we get abscissa $x_3$ of the reduced divisor $D'=P_3$ of degree $g=1$ that is enough for our purpose.
\par\noindent
\textbf{End of the reduction algorithm.}
\vskip0.2truecm

Using expressions for $A_{1,0}$ (\ref{A-coeff}), $B_{1,0}$ (\ref{B-coeff}) and $b_k$ (\ref{b-coeff}) we obtain
abscissa $x_3$ (\ref{x3-coord}) as a rational function of coordinates of points $x_{1,2}$ and $y_{1,2}$ on curve $E$. Then, substituting
\[
x_1= u1,\quad y_1= u_1\dfrac{p_{u_1}}{3}\,\qquad
x_2= u_2,\quad y_2= u_2p_{u_2}\,,
\]
we get a bulky rational function on phase space $T^*\mathbb R^2$, which we do not present here for brevity.

It is easy to verify that this function commutes with Hamiltonian $H_{31}$
\[
2H_{31}=I_1=\dfrac{u_1p_{u_1}^2}{9(u_1 - u_2)} +\dfrac{ u_2p_{u_2}^2}{u_2-u_1} -\alpha^2 (u_1^2 + u_1u_2 + u_2^2)\,,\qquad \{I_1,x_3\}=0\,,
\]
and it is functionally independent of the second St\"{a}ckel integral of motion
\[
I_2=\dfrac{u_1u_2p_{u_1}^2}{9(u_2 - u_1)} +\dfrac{ u_1u_2p_{u_2}^2}{u_1 - u_2} +\alpha^2 u_1u_2 (u_1 + u_2)\,,
\qquad \{I_2,x_3\}\neq 0\,
\]
using a modern computer algebra system.
\end{exam}

\section{Conclusion}
Integrability by quadratures is related to integration of function of the one variable. For algebraic integrals this variable can be identified with a point on the corresponding algebraic curve, that allows us to study the evolution of points along curves instead of motion in phase space. Algebraic laws determine the constants of this evolution, which can be identified with integrals of motion in phase space.

In our previous papers \cite{ts08,ts08a,ts09, ts19st,ts19r,ts19s,ts19k,ts19ar} we present some superintegrable systems with two degrees of freedom. In this note we discuss superintegrable systems with $n$-degrees of freedom for which additional integrals of motion exist due to the Riemann-Roch theorem. Our main aim is to show that such first integrals may be easily calculated using a standard reduction algorithm.

We believe that the combination of this approach based on the reduction algorithm and approach based on the bi-Hamiltonian geometry \cite{ts11} allow us to obtain additional integrals of motion without  tedious algorithmic  calculations.

The work was supported by the Russian Science Foundation (project 18-11-00032).

\end{document}